\documentclass[twocolumn,showpacs,preprintnumbers,amsmath,amssymb]{revtex4-1}
\usepackage{amsfonts}
\usepackage{amsmath}
\usepackage{graphicx}
\usepackage{dcolumn}
\usepackage{bm}
\usepackage{overpic}
\usepackage{booktabs}

\begin{document}
\title{Removing spurious interactions in complex networks}
\author{An Zeng, Giulio Cimini}
\affiliation{Department of Physics, University of Fribourg, Chemin du Mus\'{e}e 3, CH-1700 Fribourg, Switzerland}
\date{\today}

\begin{abstract}
Identifying and removing spurious links in complex networks is a meaningful problem for many real applications and is crucial for improving the reliability of network data,
which in turn can lead to a better understanding of the highly interconnected nature of various social, biological and communication systems.
In this work we study the features of different simple spurious link elimination methods, revealing that they may lead to the distortion of networks' structural and dynamical properties.
Accordingly, we propose a hybrid method which combines similarity-based index and edge-betweenness centrality.
We show that our method can effectively eliminate the spurious interactions while leaving the network connected and preserving the network's functionalities.

\end{abstract}
\maketitle

\section{Introduction}

Many social, biological and information systems are naturally described by networks, where nodes represent individuals, proteins, genes, computers, web pages, and so on,
and links denote the relations or interactions between nodes. Network analysis has hence become a crucial focus in many fields including biology, ecology, technology and
sociology~\cite{Amaral-04}. However, the reliability of network data is not always guaranteed: constructed biological and social networks may contain inaccurate and misleading information,
resulting in missing and spurious links~\cite{vonMering-02, Butts-03}.

The problem of identifying missing interactions, known as \emph{link prediction}, consists in estimating the likelihood of the existence of a link between two nodes
according to the observed links and node's attributes~\cite{Getoor-05}. Link prediction has already attracted much attention from disparate research communities due to its broad applicability.
For instance, in many biological networks (such as food webs, protein-protein interactions and metabolic networks) the discovery of interactions is often difficult and expensive,
hence accurate predictions can reduce the experimental costs and speed the pace of uncovering the truth~\cite{Clauset-08,Redner-08}. Applications in social networks include
the prediction of the actors co-starring in acts~\cite{Madadhain-05} and of the collaborations in co-authorship networks~\cite{Nowell-07}, the detection of the underground relationships
between terrorists~\cite{Clauset-08}, and many others. In addition, the process of recommending items to users can be considered as a link prediction problem in a user-item
bipartite graph~\cite{Kunegis-10}, so that similarity-based link prediction techniques have been applied to personalized recommendation~\cite{Zhang-10}.
Moreover, the link prediction approach can be used to solve the classification problem in partially labeled networks, such as predicting protein functions~\cite{Holme-05},
detecting anomalous email~\cite{Huang-06}, distinguishing the research areas of scientific publications~\cite{Gallagher-08} and finding out the fraud and legit users
in cell phone networks~\cite{Dasgupta-08}. For a review of the field, see~\cite{Lu-11}.

On the other hand, the problem of identifying spurious interactions has received less attention despite its numerous potential applications.
For instance, the identification of inactive connections in social networks or spam hyperlinks in the WWW may improve the efficiency of link-based ranking algorithms~\cite{Wang-09}, and
the detection of redundant interactions in biological, communication or citation networks may find applications in community-detection, in constructing networks' backbones~\cite{Kim-04}
or in other connection optimization problems. A possible reason for the lack of effective methods to deal with this problem is that a spurious link removal error
has far more serious consequences than a missing link addition one. If some ``unexpected'' links are incorrectly identified as spurious and removed from the network,
the system's structure and function may be altered significantly or even compromised. For instance, the network may break up into separate components so that the system's functionality is
destroyed. In power grids, only the power plants in the giant component can work~\cite{Buldyrev-10}. In traffic systems, only the cities in the giant component can mutually
communicate~\cite{Guimera-05}. In neural systems, only neurons in the giant component can reach a synchronized state and effectively process signals~\cite{Arenas-08}. The main challenge
for a spurious link detection method is hence to identify the spurious interactions and at the same time to construct a network with close functionalities to the original one.

In this work we show that many simple spurious links detection methods have indeed the serious drawback to remove real and important links, which causes the networks' structure
to be altered significantly. Hence we propose a hybrid algorithm which combines a similarity-based index known as common neighbors with the edge-betweenness centrality.
We show that this method can not only effectively identify and remove spurious links but also preserve the size of the giant component and many important structural and dynamical properties
of the network at the same time.

\section{Method}

\begin{table*}[!t]
\caption{Features of empirical networks: number of nodes ($N$) and edges ($E$), average degree ($\langle k\rangle$), average shortest path length ($\langle d\rangle$),
 clustering coefficient ($C$), degree assortativity ($r$), degree heterogeneity ($H=\langle k^2\rangle/\langle k\rangle^2$) and traffic congestability~($B_{max}$)}\label{tab.nets}
 \begin{ruledtabular}
 \begin{tabular}{lcccccccccc}
    &$N$ &$E$  &$\langle k\rangle$ &$\langle d\rangle$ &$C$ &$r$ &$H$ &$B_{max}$\\
  \hline
  CE     &$297$ &$2148$ &$14.46$ &$2.46$ &$0.308$ &$-0.163$ &$1.801$ &$2.65\cdot10^4$\\
  Email  &$1133$ &$5451$ &$9.62$ &$3.61$ &$0.220$ &$0.078$ &$1.942$ &$5.06\cdot10^4$\\
  SC     &$379$ &$914$ &$4.82$ &$4.93$ &$0.798$ &$-0.082$ &$1.663$ &$5.66\cdot10^4$\\
  PB     &$1222$ &$16717$ &$27.36$ &$2.51$ &$0.360$ &$-0.221$ &$2.970$ &$1.46\cdot10^5$\\
  PPI    &$2375$ &$11693$ &$9.85$ &$4.59$ &$0.388$ &$0.454$ &$3.476$ &$8.98\cdot10^5$\\
  USAir  &$332$ &$2126$ &$12.81$ &$2.46$ &$0.749$ &$-0.208$ &$3.464$ &$2.28\cdot10^4$
 \end{tabular}
 \end{ruledtabular}
\end{table*}

In this section we describe our procedure to study the features and evaluate the performance of a spurious link detection algorithm.
We make use of six empirical undirected networks: the C. elegans neural network (CE)~\cite{Nature393440}, an email network (Email)~\cite{PRE68065103},
a scientists' co-authorships network (SC)~\cite{PRE74036104}, the US political blogs' network (PB)~\cite{PBdata}, a protein-protein interaction network (PPI)~\cite{Nature417399}
and the US air transportation network (USAir)~\cite{USAirdata}. We only consider the giant component of these real networks. Some properties of these systems are reported in
Table~\ref{tab.nets}. All of these networks are widely used in the literature as model systems, hence we assume that they are ``true'' networks (i.e. without spurious interactions),
which we denote as $A^t$. We then add to these true networks a fraction $f$ of spurious random connections to obtain ``observed'' networks, which we denote as $A^o$,
and evaluate the ability of the spurious link detection algorithm to recover the features of the true network.

To quantify the accuracy of the algorithm in identifying the spurious interactions we use the standard metric of the area under the receiver operating characteristic curve
(AUC)~\cite{Hanley-82}. Since the algorithm returns an ordered list of links (or equivalently gives each link a score to quantify its reliability), the AUC represents the probability
that a spurious link is ranked lower than a true link. To obtain the value of the AUC, we randomly pick a spurious link and a true link in the observed network $A^o$
and compare their scores. If, among $n$ independent comparisons, the real link has higher score than the spurious link $n'$ times and equal score $n''$ times, the AUC value is:
$$\mbox{AUC}=\frac{n'+n''/2}{n}$$
Note that if links were ranked at random, the AUC value would be equal to $0.5$.

As stated in the introduction, high accuracy is not sufficient for a spurious link detection algorithm: if just a few real important links are removed,
the structural and dynamical properties of the network may change dramatically. A simple example can be seen in fig. 1. If the dashed link is removed, the network will break into
two separated components. To study the robustness of the algorithm in this respect, we remove from the observed network the fraction $f'$
of the bottom-ranked links to obtain the ``reconstructed'' network, which we denote as $A^r$. We then compare the structure and functionality of true and reconstructed networks.
We will focus mainly on giant component's (GC) size, which is of great importance for the functionality of many real systems. Then we will consider clustering coefficient~\cite{Costa-07},
average shortest path length, traffic congestability~\cite{Guimera-02} (i.e. the maximum betweenness centrality in the network) and other dynamical properties.
We will first study the case of $A^t$ and $A^r$ having the same number of links ($f'=f$). However, as in general one doesn't know how many spurious links there are in a given network,
we will finally consider the situation where $f'\neq f$.

\begin{figure}
  \center
  \includegraphics[width=6cm]{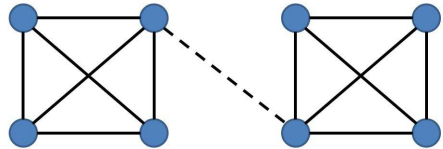}
\caption{A simple example to illustrate how an improper spurious link removal method can disconnect a network.}
\label{fig1}
\end{figure}

\section{Reliability indices}

In this section we describe some representative spurious link detection methods. These algorithms assign to each link in $A^o$ a ``reliability'' index
(denoted as $R_{ij}$ for the link connecting nodes $i$ and $j$) which quantifies the likelihood of its true existence and allows for link ranking.

\emph{Similarity-based indices} use the network's structure to assign for each pair of connected nodes $i$ $j$ a score which is directly defined as their similarity,
with the underlying assumption that a connection between similar node is likely to be a true one. These algorithms can be classified into local, quasi-local and global
according to the amount of information they need. Here are some examples:
\begin{itemize}
 \item Common Neighbors (CN): $R_{ij}^{\mbox{\tiny{CN}}}=\|\Gamma_i\cap\Gamma_j\|$,
	where $\Gamma_i$ is the set of neighbors of node $i$ and $\|\ldots\|$ indicates the number of nodes in a set.
 \item Resource Allocation (RA): $R_{ij}^{\mbox{\tiny{RA}}}=\sum_{k\in\Gamma_i\cap\Gamma_j}\frac{1}{\|\Gamma_k\|}$.
 \item Local Path (LP): $R_{ij}^{\mbox{\tiny{LP}}}=(A^2)_{ij}+\epsilon\:(A^3)_{ij}$, where $A$ is the network's adjacency matrix and $\epsilon<1$ is a free parameter.
 \item Katz index (Katz): $R_{ij}^{\mbox{\tiny{Katz}}}=\sum_{l=1}^{\infty}[(\beta\,A)^l]_{ij}$, where $\beta$ is a free parameter which must be lower than the reciprocal
       of the largest eigenvalue of $A$.
\end{itemize}

\emph{Centrality-based indices} measure the importance of a link in the network, assuming that the higher the link's centrality, the higher its reliability.
We consider two simple indices:
\begin{itemize}
\item Preferential Attachment (PA): $R_{ij}^{\mbox{\tiny{PA}}}=\|\Gamma_i\|\times\|\Gamma_j\|$.
\item Edge Betweeness (EB): $R_{ij}^{\mbox{\tiny{EB}}}=\sum_{m>n}\frac{C_{mn}^{(ij)}}{C_{mn}}$, where $C_{mn}$
      is the number of shortest paths from node $m$ to node $n$ and $C_{mn}^{(ij)}$ is the number of such shortest paths passing through the link $ij$.
\end{itemize}

Clearly, CN, RA and PA are local indices. CN is the simplest possible measure of neighborhoods' overlap, while RA \cite{Zhou-09} is the best performing local index
for the purpose of link prediction. PA is the algorithm which requires less information. LP \cite{Zhou-09} is instead a quasi-local method,
as it considers local paths with wider horizon than CN (it also counts the number of different paths with length $3$ connecting $i$ and $j$).
Finally, Katz \cite{Kats-53} and EB methods are global indices, as they are based on the ensemble of all paths in the network.
Specifically, Katz counts the paths between two nodes and weights them according to their length $l$,
while EB is built with the number of shortest paths from all vertices to all others that pass through the given link.

\section{Hybrid index}

We now introduce a hybrid index which combines the similarity-based and the centrality-based approaches.
The underlying idea is that we consider a link to be a ``true'' one either if it connects similar nodes or if it has a central position in the network.
Even if this assumption is not necessarily true, as we will show later it avoids the removal of important links so that the network's properties and functions are preserved,
with the small drawback of failing to identify few spurious interactions.

To construct the Hybrid index, we combine the simple common neighbor with edge-betweenness centrality as:
$$R_{ij}^{\mbox{\tiny{hyb}}}=\lambda\:\frac{R_{ij}^{\mbox{\tiny{CN}}}}{\mbox{max}_{mn}(R_{mn}^{\mbox{\tiny{CN}}})}
+(1-\lambda)\:\frac{R_{ij}^{\mbox{\tiny{EB}}}}{\mbox{max}_{mn}(R_{mn}^{\mbox{\tiny{EB}}})}$$
where $\lambda\in[0,1]$ is the hybridization parameter. In what follows we set $\lambda=0.9$, because we want to exploit mainly CN
and a small contribution from EB will suffice for our purposes (however, see section \ref{sec.hyb} for a study of the index behavior for different $\lambda$).
Note that this is only one possibility of defining such index. We made use of CN because it is the most well-known of the similarity-based indices.
However one could use e.g. RA or Katz instead, though the qualitative features of the Hybrid method wouldn't change.

\section{Results}

\begin{figure}
  \center
  \includegraphics[width=9cm]{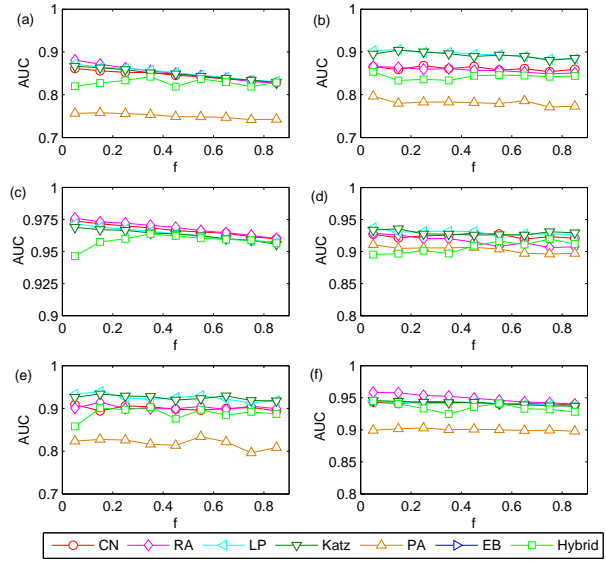}
\caption{(Color online) AUC for various indices and for different values of $f$. The true networks are (a)CE, (b)Email, (c)SC, (d)PB, (e)PPI, (f)USAir.
Results are averaged over $100$ independent realizations. Note that the curves for EB are not shown as the respective AUC values are too low. The same holds for PA in panel (c).}
\label{fig2}
\end{figure}

In this section we compare the features of the spurious link detection approaches which have been previously introduced.
We start by adding to the true networks $A^t$ a fraction $f$ of random connections to obtain the observed networks $A^o$. For each particular index,
we rank the links according to their reliability values and measure the accuracy of the method in identifying spurious interactions by the AUC (Figure \ref{fig2}).
We observe that generally the similarity-based methods perform better than the centrality-based ones. Among the first category, Katz and LP \cite{foot1} perform slightly better than CN and RA
as they take advantage of using more information. Among the second, EB is the worst performing, with AUC even lower than 0.5.
The performance of the Hybrid method is instead very close to that of the pure similarity-based indices. Hence having a contribution from EB in the hybridization
does not result in worse spurious link detection (as one might expect).

We already argued that accuracy is not the only criterion to assess the performance of these methods.
The other important aspect is that the removal of putative spurious links should not alter the giant component's size as well as other properties of the networks.
To investigate this aspect, we remove from $A^o$ the fraction $f'$ of the bottom-ranked links to obtain the reconstructed networks $A^r$, whose features we compare with the ones
of the relative true networks $A^t$. We start with the simple case $f'=f$ and we first focus on the GC's size, which is of great relevance in many contexts.
\begin{figure}
  \center
  \includegraphics[width=9cm]{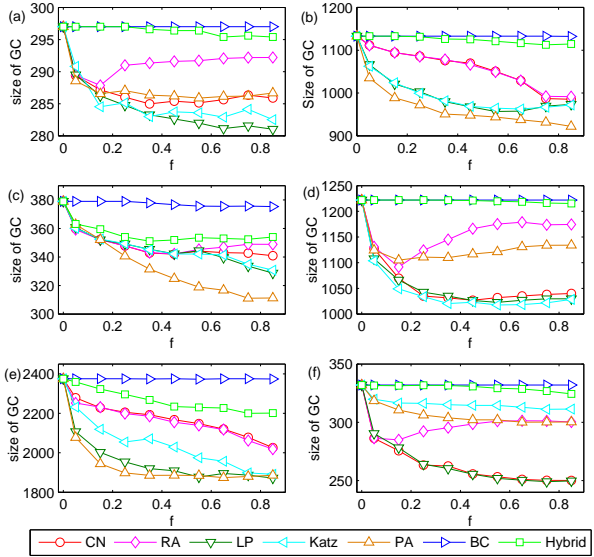}
\caption{(Color online) GC's size when various indices are used to build $A^r$ (here $f'=f$) and for different $f$.
The true networks are (a)CE, (b)Email, (c)SC, (d)PB, (e)PPI, (f)USAir. Results are averaged over $100$ independent realizations.}
\label{fig3}
\end{figure}
As shown in Figure \ref{fig3}, the GC's size significantly decreases with $f'$ when using any similarity-based method (as well as PA): in these cases
many nodes becomes disconnected from the networks' core and end up losing their function. On the contrary, EB always keeps the networks connected.
This is not surprising, as it has already been pointed out~\cite{PR48675} that similarity indices and EB are highly anti-correlated,
meaning that removing links between non-similar nodes causes links with high betweenness to be cut, and vice-versa.
What is remarkable is that also the Hybrid method can effectively preserve the connectedness of the original networks in most of the cases,
and in general much better than any other similarity-based method, despite the small contribution it receives from EB.
It is hence sufficient to increase little the reliability of central and important links to avoid removing them.

\begin{figure}
  \center
  \includegraphics[width=9cm]{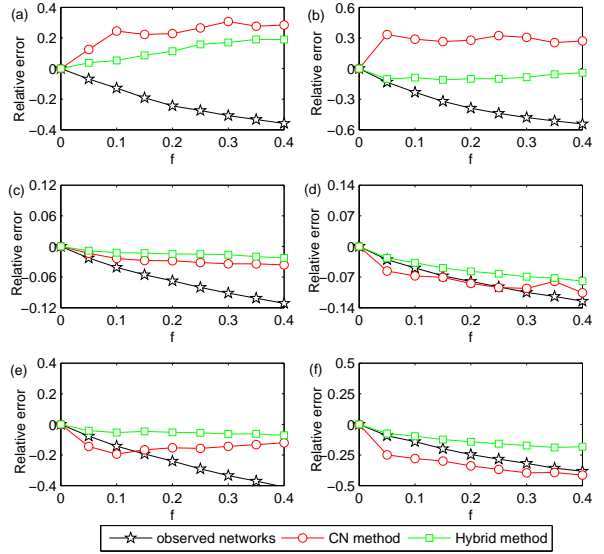}
\caption{(Color online) Relative errors of clustering coefficient (a)-(b), average shortest path length (c)-(d) and transportation congestability (e)-(f) for different $f$.
The different lines correspond to the relative errors in $A^o$ and in the two $A^r$ built by CN and Hybrid methods respectively, with $f'=f$.
Left plots refer to PB while right plots to USAir. Results are averaged over $100$ independent realizations.}
\label{fig4}
\end{figure}

We move further by considering other network properties. In order to compare the true and the reconstructed networks under a given property $X$,
we compute the relative error of $X$ as $(X(A^r)-X(A^t))/X(A^t)$. As a benchmark, we also compute the relative error of $X$ in the observed networks as $(X(A^o)-X(A^t))/X(A^t)$.
For an effective spurious link removal method, which is able to reproduce the properties of the true network, the absolute value of the relative error for $A^r$ should be smaller
than the absolute value of the relative error for $A^o$ (meaning that $A^r$ is a better estimate of $A^t$ than $A^o$)
and as close as possible to zero (meaning that $X$ has approximately the same value in $A^t$ and $A^r$).
Figure \ref{fig4} shows the relative errors made by CN and Hybrid methods for clustering coefficient, average shortest path length and traffic congestability
(i.e. the maximum betweenness centrality in the network). We only report the results for the Political blog (PB) and US Airline (USAir) networks,
as these are the cases in which the GC's size is relatively more affected when using pure similarity-based methods (Figure \ref{fig3}).
We observe that in these cases the Hybrid method is always able to restore the properties of the true network with respect to the observations, while this is not always true for CN.
Moreover, the Hybrid method always preserves the networks' properties better than CN, at the small cost of achieving smaller AUC values.
This is because CN and other similarity-based methods alter the GC, which is much more harmful for the networks' properties and functions than keeping fewer more spurious links.
Note however that if the CN method does not cause serious enough damage to the GC---as it happens for C. elegans neural (CE) and scientists' co-authorships (SC) networks---then
the situation may be reversed: CN can preserve some of the network properties better than the Hybrid method due to its higher accuracy.

There are plenty of other network's static and dynamical properties which can be considered, such as synchronization, spreading threshold, and so on.
As these dynamics can only take place in the GC, similarity-based methods which break the network into pieces alter them seriously.
For example, the nodes out of the GC can never reach the global synchronized state, and the signal from the GC can never spread to these nodes.
Again, these methods eventually destroy the system's functions.

As in real applications of spurious links removal one does not know the exact number of spurious links in a network, we finally consider the case when $f'\neq f$.
To do so, we fix the number of random connections added to $A^t$ at $f=10\%$. We then study the properties of the networks $A^r$ reconstructed by different methods
by removing different fractions $f'$ of links from $A^o$.
\begin{figure}
  \center
  \includegraphics[width=9cm]{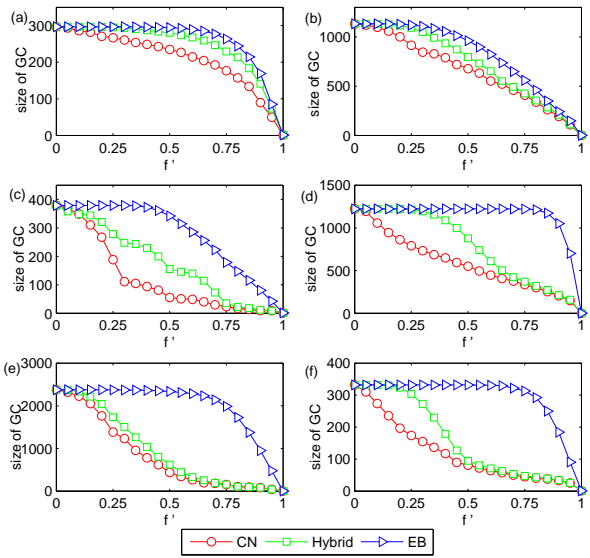}
\caption{(Color online) The GC's size when different fractions of links $f'$ are removed from $A^o$ by CN, Hybrid and EB methods.
The true networks are (a)CE, (b)Email, (c)SC, (d)PB, (e)PPI, (f)USAir. Results are averaged over $100$ independent realizations.}
\label{fig5}
\end{figure}

\begin{figure}
  \center
  \includegraphics[width=9cm]{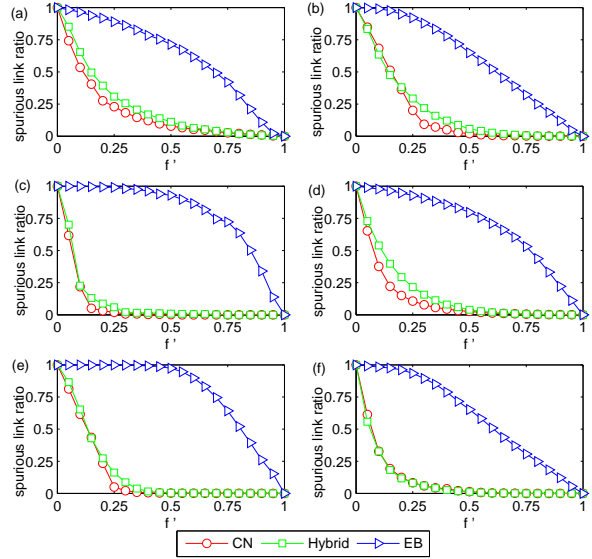}
\caption{(Color online) The residual fraction of spurious links in $A^r$ when different fractions of links $f'$ are removed from $A^o$ by CN, Hybrid and EB methods.
The true networks are (a)CE, (b)Email, (c)SC, (d)PB, (e)PPI, (f)USAir. Results are averaged over $100$ independent realizations.}
\label{fig6}
\end{figure}

Figure \ref{fig5} shows the GC's size for varying $f'$.
We observe that the GC's size naturally decreases with the fraction of removed links. Such decrease is very fast when using CN and very slow when using EB---in the latter case,
the GC's size is preserved in any network even when half of the links are removed. The Hybrid method lies between these two, and remarkably it performs like EB
when the fraction of removed links is not too big (in many cases the GC's size has a plateau which may last up to large $f'$).
Another interesting aspect would be to investigate how many of the original $f$ spurious links are left in the networks for various $f'$. Results are shown in Figure~\ref{fig6}.
We again observe that the more we remove links, the higher the probability to remove a spurious link.
Due to its low accuracy, EB must remove almost all links in order to get rid of the spurious ones. On the contrary, CN can eliminate all the spurious links quite soon ($f'\simeq 25\%$).
Interestingly, the Hybrid performs as well as CN and their curves almost overlap.
These results again indicate that the Hybrid method represents an effective approach to both preserve the GC's size and to achieve high accuracy.
Moreover, it is also more robust than other methods when considering the intrinsic uncertainty of the number of spurious interactions in a system.

\section{The Hybridization parameter}\label{sec.hyb}

\begin{figure}
  \center
  \includegraphics[width=9cm]{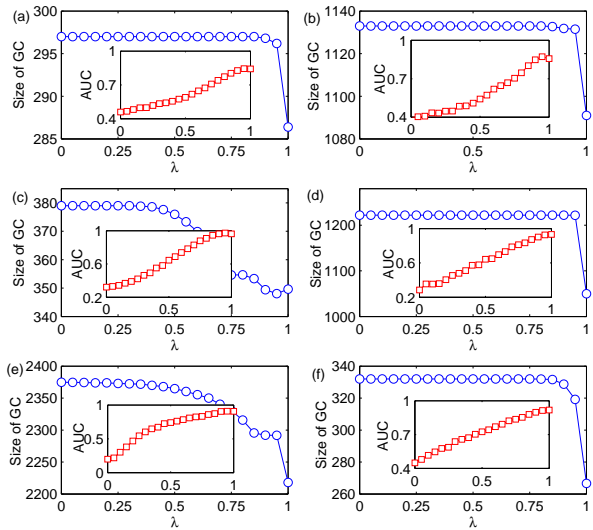}
\caption{(Color online) The size of the GC in the networks reconstructed by the Hybrid method with different values of $\lambda$.
Insets: the AUC for different $\lambda$. The respective true networks are (a)CE, (b)Email, (c)SC, (d)PB, (e)PPI, (f)USAir.
Results are averaged over $100$ independent realizations.}
\label{fig7}
\end{figure}

At last, we show how the Hybrid index behaves by varying the value of the parameter $\lambda$. In order to do so, we consider the particular case in which the observed networks $A^o$
are obtained from the true networks $A^t$ with the addition of $f=20\%$ of spurious links. Figure~\ref{fig7} shows AUC and GC'size of the networks $A^r$ reconstructed
by the Hybrid method (with $f'=f$) for different values of $\lambda$. We observe that while the AUC decreases for decreasing $\lambda$
(but this decrease is always slower at the beginning), the GC remains almost integer except when $\lambda$ becomes too close to 1.
Therefore it is sufficient to have a small contribution from EB in the Hybrid method to keep the network connected at the cost of being slightly less accurate.
This is the reason why we have previously set $\lambda=0.9$. Note that one can always use a bigger value of $\lambda$ if accuracy is the main goal,
or a smaller value if the GC's integrity is a major issue.

\section{Discussion}

How to detect and remove spurious interactions in networks is a significant problem which may find application in almost any field of complex science.
Still, it has not yet attracted much attention, as the consequences of a removal error can heavily harm the system under investigation.
In the literature many similarity-based methods for the purpose of link prediction have been proposed. In this work we showed that, when applied to spurious link detection,
all these methods achieve high accuracy but suffer from the important drawback of decreasing the size of the giant component
and distorting other static and dynamic properties of the network. This harmful effect may cause a system to lose its functions,
as nodes which are disconnected from the GC cannot communicate with the network's core.
In order to overcome these drawbacks, we proposed a hybrid method which combines the similarity-based common neighbors index with edge-betweenness centrality.
We showed that this approach can effectively eliminate the spurious links and at the same time keep the network connected; moreover important properties like
clustering coefficient, average shortest path length and traffic congestability can be generally preserved better.
This method is still more advantageous when the number of spurious interactions within a system is unknown.

In the literature there are other important examples of spurious link detection approaches (e.g. hierarchical random graph \cite{Clauset-08}
and stochastic block model~\cite{Guimera-09}) which however were not focusing on preserving the giant component's size.
Moreover these methods are based on global algorithms which can be prohibitive to use for large-scale systems.
Our method instead would be easily applicable for large networks. This is because it combines common neighbors index, which requires only local information of a link,
and edge-betweenness centrality, whose computational complexity is now as lower as $O(NE)$, where $N$ and $E$ are respectively the number of nodes and edges in the network~\cite{JMS25163}.

Finally, we remark that the problem of identifying spurious interactions is much more difficult to deal with than predicting missing interactions.
We already pointed out how serious a removal error may be. In addition, while in link prediction studies there's a true network from which some existing links are removed
to generate the observation and test the algorithm, for spurious link detection how to add spurious interactions to the true network is generally unknown.
In this work we explored the simplest situation, in which spurious links are just random connections between nodes. This approach can be suitable for describing some systems
(for instance biological networks obtained from measurements prone to random errors, or social networks in which some links result from once in a lifetime interactions between people)
but may result inadequate for others (like biological systems when measurements are prone to systematic errors, or the WWW where spam hyperlinks always start from the same set of pages).
The effectiveness of a spurious link detection method in these systems hence deserve further validation, which will be the subject of future work.

\section*{Acknowledgments}
We would like to thank Yi-Cheng Zhang, Mat\'{u}\v{s} Medo, Chi Ho Yeung and Stanislao Gualdi for helpful suggestions. 
This work is partially supported by the Swiss National Science Foundation under Grant No. 200020-132253 
and by the Future and Emerging Technologies program of the European Commission FP7-COSI-ICT (project QLectives, grant no. 231200).


\begin{thebibliography}{99}
\bibitem{foot1} For LP and Katz we set the parameters $\epsilon$ and $\beta$ to the values which maximize the respective AUC.
\end{thebibliography}
\end{document}